\documentclass[journal,twoside,web]{ieeecolor}
\usepackage{tabu}
\usepackage{jsen}
\usepackage{cite}
\usepackage{etoolbox}
\usepackage{amsmath,amssymb,amsfonts}
\usepackage{algorithmic}
\usepackage{graphicx}
\usepackage{textcomp}
\usepackage{wrapfig}
\usepackage{siunitx}
\usepackage{booktabs} 
\makeatletter
\def\@IEEEtriggercmd{\enlargethispage{-1in}}
\makeatother

\def\BibTeX{{\rm B\kern-.05em{\sc i\kern-.025em b}\kern-.08em
    T\kern-.1667em\lower.7ex\hbox{E}\kern-.125emX}}
\definecolor{abstractbg}{rgb}{0.89804,0.94510,0.83137}
\begin{document}

\title{Bioinspired Tapered-Spring Turbulence Sensor for Underwater Flow Detection}
\author{Xiao Jin, Zhenhua Yu*, Thrishantha Nanayakkara
\thanks{
*This research is partially supported by the European Union’s Horizon 2020 Research and Innovation Programme under Grant Agreement No. 101016970 (Natural Intelligence).}
\thanks{Xiao Jin, Zhenhua Yu and Thrishantha Nanayakkara are with Dyson School of Design Engineering, Imperial College London, SW7 2DB, London, UK, and Zhenhua Yu is also with Department of Computer Science, University of Aberdeen, AB24 3UE, Aberdeen, UK}
}

\IEEEtitleabstractindextext{%
\fcolorbox{abstractbg}{abstractbg}{%
\begin{minipage}{\textwidth}%
\begin{wrapfigure}[16]{r}{3in}%
\includegraphics[width=2.9in]{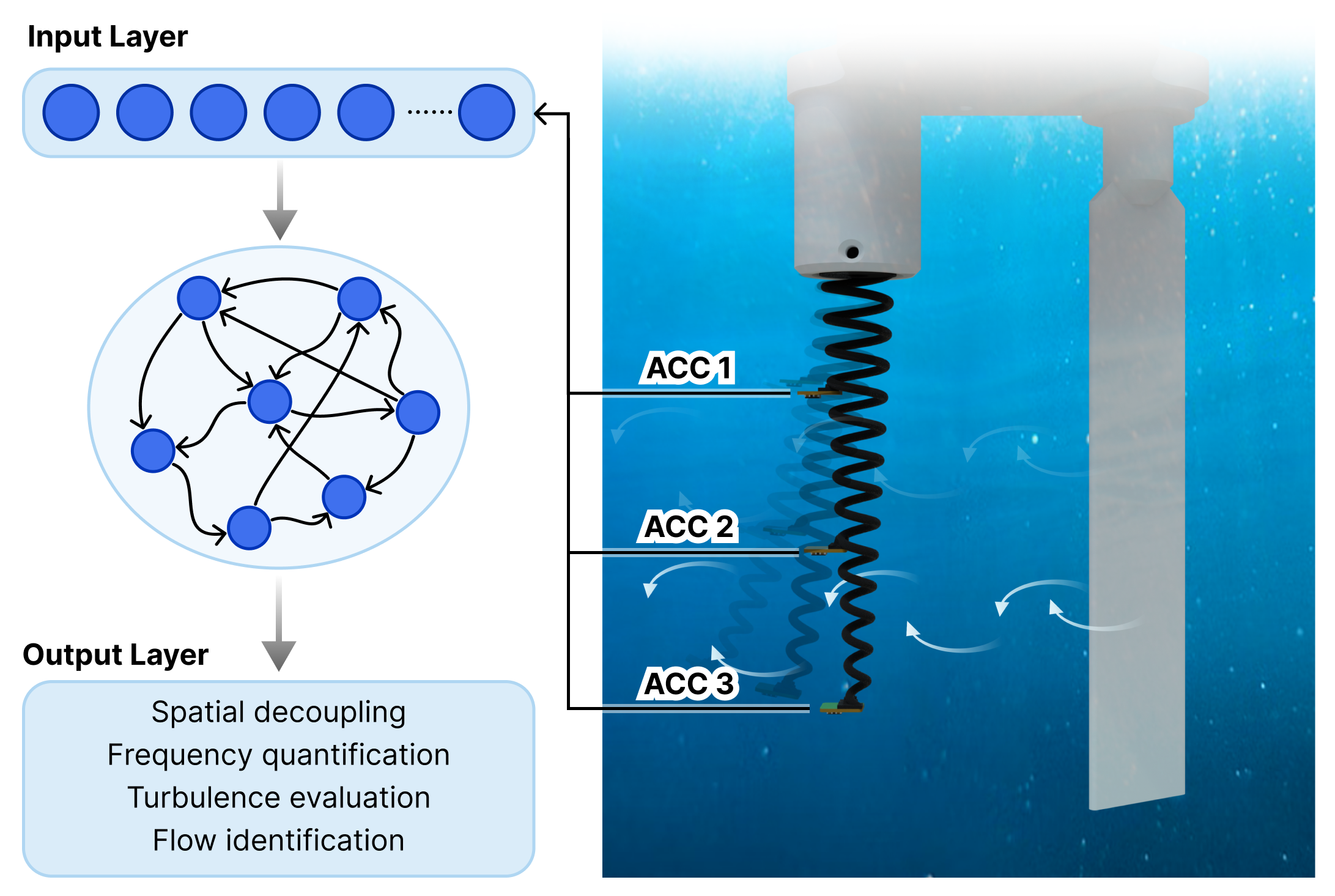}%
\end{wrapfigure}%
\begin{abstract}

This paper presents a bio-inspired underwater whisker sensor for robust hydrodynamic disturbance detection and efficient signal analysis based on Physical Reservoir Computing (PRC). The design uses a tapered nylon spring with embedded accelerometers to achieve spatially distributed vibration sensing and frequency separation along the whisker. Towing-tank experiments and computational fluid dynamics simulations confirmed that the whisker effectively distinguishes vortex regimes across different fin angles and maintains Strouhal scaling with flow velocity, where higher speeds increase vibration intensity without affecting the dominant frequencies. Frequency-domain analysis, Shannon entropy, and machine learning further validated the sensing performance: vortex shedding frequencies were identified with less than 10\% error, entropy captured the transition from coherent vortex streets to turbulence, and logistic regression achieved 89.47\% classification accuracy with millisecond-level inference. These results demonstrate that structurally encoded whisker sensing provides a scalable and real-time solution for underwater perception, wake tracking, and turbulence-aware navigation in autonomous marine robots.

\end{abstract}

\end{minipage}}}

\maketitle

\section{Introduction}
\label{introduction}
Underwater environments, ranging from shallow coastal zones to the deep ocean floor, remain some of the least explored areas on Earth\cite{NAP10844} \cite{water}. These environments are not only ecologically vital, hosting complex marine ecosystems\cite{bg-7-2851-2010}, but also hold significant economic and scientific value due to their potential for resource extraction\cite{LEARY2009183}, infrastructure development, and climate monitoring. However, underwater exploration faces unique challenges such as limited visibility\cite{6410440}, strong current disturbances, and complex terrain geometries, which constrain the effectiveness of conventional sensing technologies like vision-based systems or sonar\cite{4515907}. As the demand grows for autonomous underwater vehicles (AUVs)\cite{RIDAO2015227} capable of precise and reliable perception in these conditions, there is a pressing need to develop alternative sensing strategies that can operate robustly under turbid, dynamic, and unstructured environments.

To meet the increasing demands of underwater exploration, various types of autonomous underwater vehicles (AUVs) have been developed \cite{jmse11061119,ALAM201416}, ranging from compact observation-class platforms to large intervention-class systems. These robots are widely deployed for applications such as oil and gas exploration \cite{gilmour2012field}, infrastructure inspection, environmental monitoring \cite{doya2017seasonal,laschi2016soft}, pipeline tracking \cite{jawhar2018architecture}, biological sampling \cite{yoshida2007deepest}, and target retrieval. In both industrial and scientific contexts, AUVs are expected to operate autonomously for extended durations, navigate through unstructured and dynamic underwater terrains, and perform complex tasks with limited communication and sensing capabilities. Although significant progress has been made in navigation and control, most AUVs still rely heavily on sonar \cite{sonar,sonar2}, doppler velocity logs (DVL) \cite{teixeira2016auv}, and inertial navigation systems (INS) \cite{jalving2003toolbox} for environmental perception. However, these conventional sensors often struggle in cluttered, turbid, or highly dynamic environments where visual occlusion \cite{water1,4524846}, acoustic scattering, and current-induced drift are prevalent. As a result, AUVs frequently fail to detect fine-grained terrain features, deformable or flexible obstacles, and subtle environmental cues. More critically, they remain largely blind to essential hydrodynamic information such as localized flow disturbances, shear layers, vortex streets, and small-scale eddies—features that often contain crucial information about nearby obstacles, terrain boundaries, or biological activity. The inability to capture such near-field flow dynamics severely limits an AUV's ability to adapt and interact effectively with its surroundings in real time. This highlights the urgent need for advanced sensing mechanisms capable of resolving subtle variations in flow direction, velocity, and turbulence—capabilities that lie beyond the reach of traditional sonar- and inertial-based systems.

To overcome these limitations, a shift is needed from traditional passive and remote sensing toward embodied, contact-based strategies that emulate biological systems\cite{wang2023potential,xia2023current}. In nature, certain aquatic mammals have evolved highly sensitive tactile sensors—whiskers—that enable them to detect minute flow disturbances, track hydrodynamic trails, and interact effectively with their surroundings in complete darkness\cite{w1,10907784}. Inspired by the extraordinary hydrodynamic sensitivity of harbor seal whiskers, several researchers have investigated bioinspired flow sensors that mimic their morphology and function.  Beem et al.\cite{Beem_2013} developed and validated a seal whisker-inspired sensor capable of detecting low-amplitude water disturbances that are typically undetectable by conventional flow sensors.  Kottapalli et al.\cite{kottapalli2015seal} proposed a MEMS-based flow sensor designed to reduce vortex-induced vibrations by replicating the undulated shape of seal whiskers.  Asadnia et al.\cite{Asadnia} introduced artificial fish skin embedded with micro-hair cells to sense subtle flow variations, extending the biological principles of whisker sensing to broader underwater contexts.  These efforts highlight the potential of whisker-based designs in capturing fine-scale hydrodynamic cues for underwater robotic applications.

Despite significant progress in bioinspired underwater whisker sensors, most existing designs still suffer from high power consumption, limited frequency–spatial decoupling capability, and heavy reliance on computational post-processing for feature extraction. These limitations restrict their applicability in low-power autonomous underwater platforms that require real-time hydrodynamic perception.

To address these gaps, we present a tapered-spring-based whisker integrated with Physical Reservoir Computing (PRC) and three spatially distributed MEMS IMU nodes. This design passively encodes flow-induced oscillations into multi-modal mechanical states, enabling intrinsic frequency–spatial separation along the whisker length. Coupled with lightweight feature extraction and logistic regression classification, our approach achieves sub-4\% vortex frequency extraction error and millisecond-level inference latency with minimal power consumption, making it suitable for long-duration deployments on resource-constrained AUVs.   Compared to existing underwater whisker sensors, our design offers ultra-low power consumption, intrinsic frequency–spatial decoupling, and millisecond-level inference latency using lightweight machine learning models.

Compared to existing underwater whisker sensors, our design offers the following key advantages:
\begin{enumerate}
    \item \textbf{Ultra-low power consumption} – A custom PCB integrating TDK ICM-42670-P 6-axis IMUs operates at \SI{3.0}{\volt}, consuming only \SI{68}{\micro\ampere} (accelerometer) and \SI{370}{\micro\ampere} (gyroscope) at \SI{50}{\hertz}, corresponding to a total power draw of approximately \SI{1.3}{\milli\watt}, enabling long-term deployment without active power refresh.
    \item \textbf{Intrinsic frequency–spatial decoupling} – The tapered-spring geometry and three distributed sensing nodes passively separate modal responses along the whisker length, eliminating the need for heavy digital filtering.
    \item \textbf{Lightweight real-time processing} – Logistic regression classification achieves ${85.4\%}$ accuracy with $<$\SI{0.001}{ms} inference latency, making onboard real-time classification feasible for small-scale AUVs.
    \item \textbf{Robust underwater performance} – The design maintains $<{10\%}$ vortex frequency extraction error despite hydrodynamic damping, as verified through towing tank tests and CFD-based harmonic analysis.
\end{enumerate}

The remainder of this paper is organized as follows. Section \ref{s} details the design of the bio-inspired whisker sensor, including the tapered-spring geometry, mechanical modeling, and embedded accelerometer configuration. Section \ref{result} describes the experimental setup, towing-tank validation, and corresponding computational fluid dynamics simulations, followed by analysis of frequency responses, entropy-based flow characterization, and logistic-regression-based classification.

\section{Sensor Design and Methods}
\label{s}

To enable robust perception in fluid environments, we propose a bio-inspired whisker sensor that integrates principles of physical reservoir computing (PRC) with a tapered spring structure. PRC has recently emerged as a promising framework for temporal signal processing, leveraging the intrinsic dynamics of physical substrates to transform input stimuli into high-dimensional representations without requiring complex digital computation\cite{Hauser2011}. In whisker-based sensing, this principle can be embodied through structural elements that respond to environmental forces with rich, nonlinear dynamics\cite{Yu2023}.

Inspired by the conical morphology of mammalian vibrissae, the proposed whisker adopts a tapered helical spring as its core mechanical structure\cite{yu}. This geometry, widely observed in terrestrial rodents, enables graded stiffness along the length of the whisker and promotes spatially distributed oscillations when interacting with external stimuli. The gradual tapering facilitates both localized bending near the tip and stable anchoring at the base, offering favorable conditions for encoding dynamic input into spatiotemporal mechanical states.
\subsection{Design of Bio-Inspired Whisker}

In underwater conditions, complex flow phenomena such as Kármán vortex streets introduce nonstationary and nonlinear disturbances with distinct frequency structures. Instead of explicitly extracting these features through computational filtering or spectral analysis, our approach leverages the PRC paradigm to passively encode them into the vibrational dynamics of the whisker itself. The tapered structure naturally transforms flow-induced vortex shedding into spatially and temporally differentiated mechanical states, which are recorded by embedded IMU sensors. This structural computation allows the system to recognize characteristic hydrodynamic signatures—such as vortex shedding frequency and wake asymmetry—without requiring active sensing or external processing, providing a lightweight and efficient solution for real-time underwater perception.

\begin{figure}[ht!] 
\centering
\includegraphics[width=3.3in]{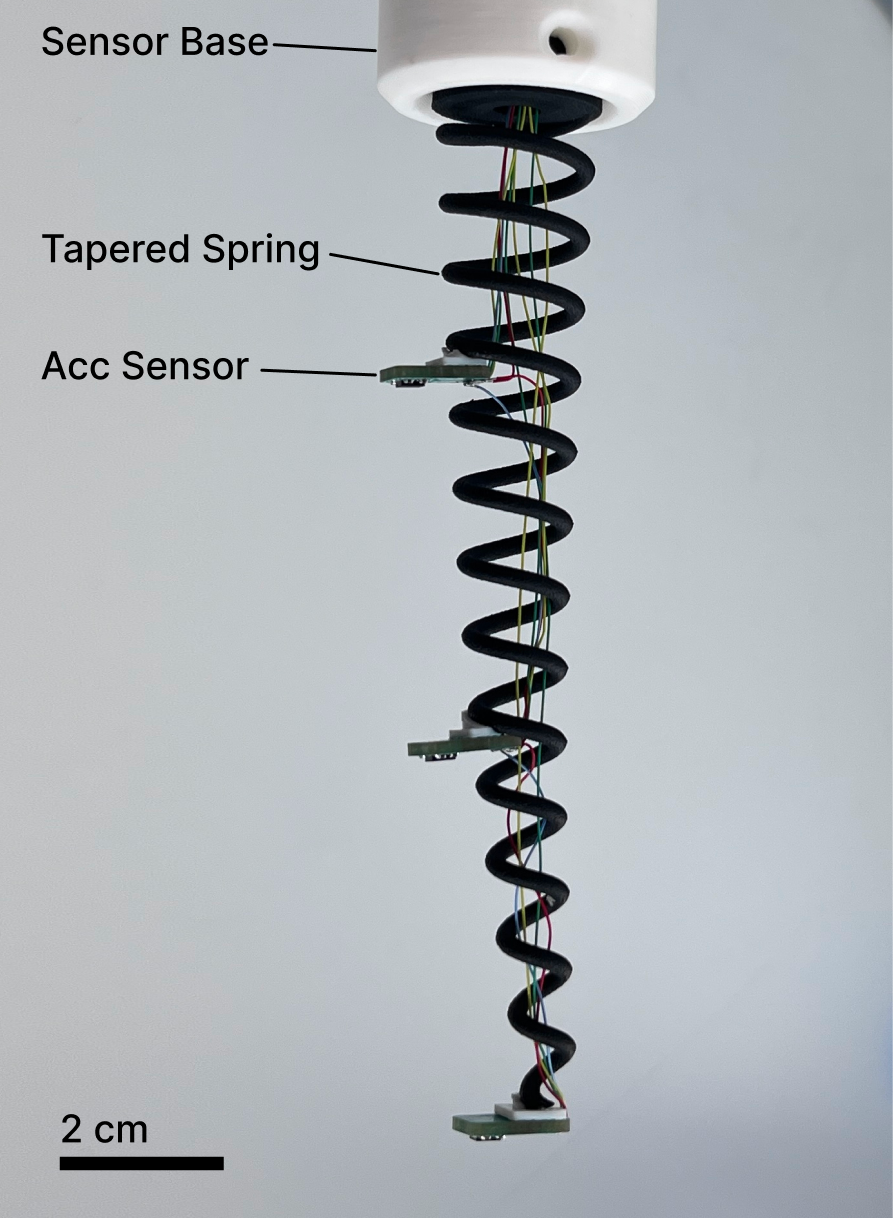}
\caption{
Tapered spring-based whisker sensor for fluid interaction sensing. The system includes a 3D-printed nylon PA12 conical spring with embedded accelerometers, enabling spatially distributed vibration sensing for physical reservoir computing.
}
\label{se}
\end{figure}

While the tapered spring alone can exhibit nonlinear dynamic responses suitable for physical reservoir computing in air, its underwater deployment requires additional structural considerations as shown in Figure.\ref{se}. The core structure is fabricated using resin-based additive manufacturing, forming a conical spring with a total height of 122mm, a base outer diameter of 22mm, and a tip diameter of 7.2mm. The spring wire has a uniform diameter of 2.6mm, chosen to balance stiffness and compliance along the length of the whisker.

To ensure robust mechanical performance and preserve the rich nonlinear dynamics required for physical reservoir computing (PRC), the whisker is fabricated from high-strength nylon (PA12) using selective laser sintering (SLS). This material offers a flexural strength of approximately 90MPa and provides a favorable balance between structural compliance and environmental durability, making it well-suited for both air and underwater deployment without requiring additional encapsulation.

Rather than emulating the external morphology of biological vibrissae, our design adopts a mechanism-inspired approach that leverages the tapered stiffness gradient of a helical spring to physically amplify subtle environmental disturbances. This structural configuration enables spatially distributed deformation, encoding dynamic inputs into rich mechanical responses—functionally analogous to how mammalian whiskers convert external forces into neural signals via localized bending and tension along the follicle.

In engineered contexts, however, such neural processing is often impractical or costly to replicate. The proposed structure addresses this challenge by embedding signal processing capabilities directly into the mechanics of the system. The conical spring geometry transforms spatially varying forces—such as those from fluid interactions or mechanical contacts—into rich, distributed oscillatory responses. These mechanical states, shaped by the graded stiffness and inertial properties along the whisker, serve as a physical substrate for temporal signal encoding without digital preprocessing. By replacing complex sensor arrays and high-bandwidth computation with a materially-encoded reservoir, this design enables low-cost, robust, and energy-efficient sensing in dynamic environments.

\subsection{Experimental Setup for Fluid-Stimulated Vibration Acquisition}
\color{black}
To validate the persistence of physical reservoir computing (PRC) characteristics under fluid-induced excitation, a series of computational and numerical analyses were conducted in an underwater environment. The bio-inspired whisker sensor, previously described as a tapered spring structure encapsulated by a PVC membrane, was subjected to harmonic response analysis to evaluate its ability to preserve spatially distributed resonance behavior under realistic flow conditions.

\begin{figure}[t]
    \centering
    \includegraphics[width=0.48\textwidth]{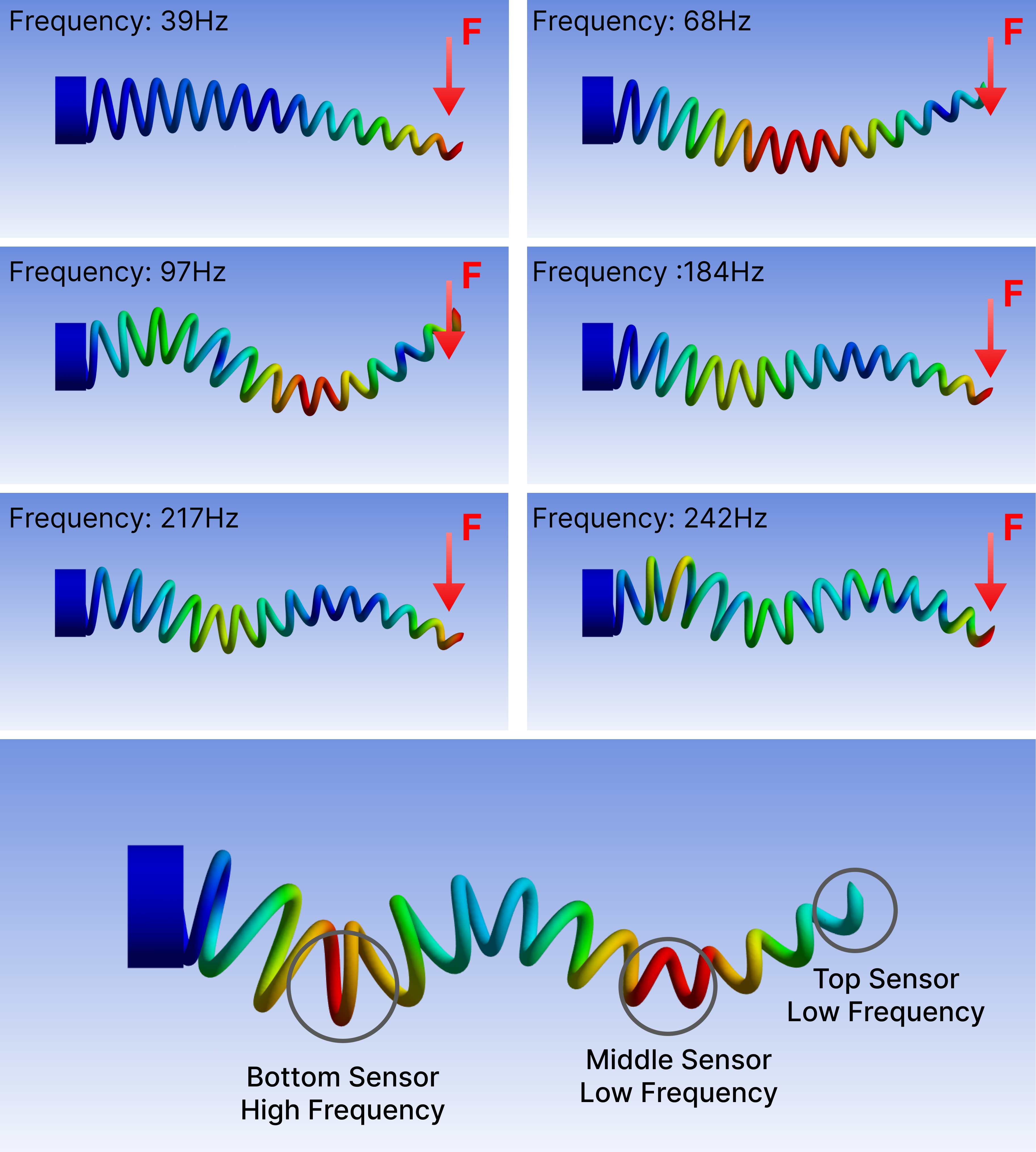}
    \caption{Harmonic response modes of the encapsulated whisker structure at various frequencies. Three distinct modal concentrations emerge along the length, enabling spatial encoding under fluid-induced oscillation.}
    \label{fig:modal}
\end{figure}

To approximate the hydrodynamic loading exerted by environmental fluid flows, the projected frontal area of the whisker was modeled as a trapezoidal profile with a top width of 7.2\,mm, a base width of 22\,mm, and a height of 122\,mm, yielding an effective area of approximately 0.00177\,m\textsuperscript{2}. Using the standard drag force equation:

\begin{equation}
F_D = \frac{1}{2} \rho C_D A v^2,
\end{equation}

where $\rho = 1000\,\mathrm{kg/m^3}$ is the density of water, $C_D = 1.2$ is the assumed drag coefficient for a conical or tapered bluff body, and $v$ is the local flow velocity, the expected drag forces were estimated. Depending on flow speed, the drag ranged from approximately 0.0106\,N (at 0.1\,m/s) to 0.266\,N (at 0.5\,m/s). These values represent realistic fluid stimuli that the whisker structure may encounter in underwater environments.

To simulate a conservative excitation condition and ensure the activation of a full set of modal responses, a lateral harmonic force was applied to the tip of the structure in the finite element model. This excitation magnitude exceeds the expected upper bound of natural fluid drag, ensuring the observability of the vibrational characteristics across the entire operational spectrum. As shown in Fig.~\ref{fig:modal}, six dominant resonant modes were identified at 39\,Hz, 68\,Hz, 126\,Hz, 184\,Hz, 217\,Hz, and 242\,Hz. Each modal shape exhibits spatially localized deformation patterns along the whisker body, corresponding to distinct vibrational nodes and antinodes.

Notably, three particularly prominent modal concentrations can be observed, suggesting that the system inherently supports distributed encoding of external stimuli across both frequency and spatial domains. This multi-modal behavior, resilient to structural encapsulation, confirms the design's capacity to function as a physical reservoir under submerged conditions, enabling PRC-based signal transformation and flow sensing.

\subsection{Accelerometer Sensor Design and Installation}

\color{black}

\begin{figure}[t]
    \centering
    \includegraphics[width=0.47\textwidth]{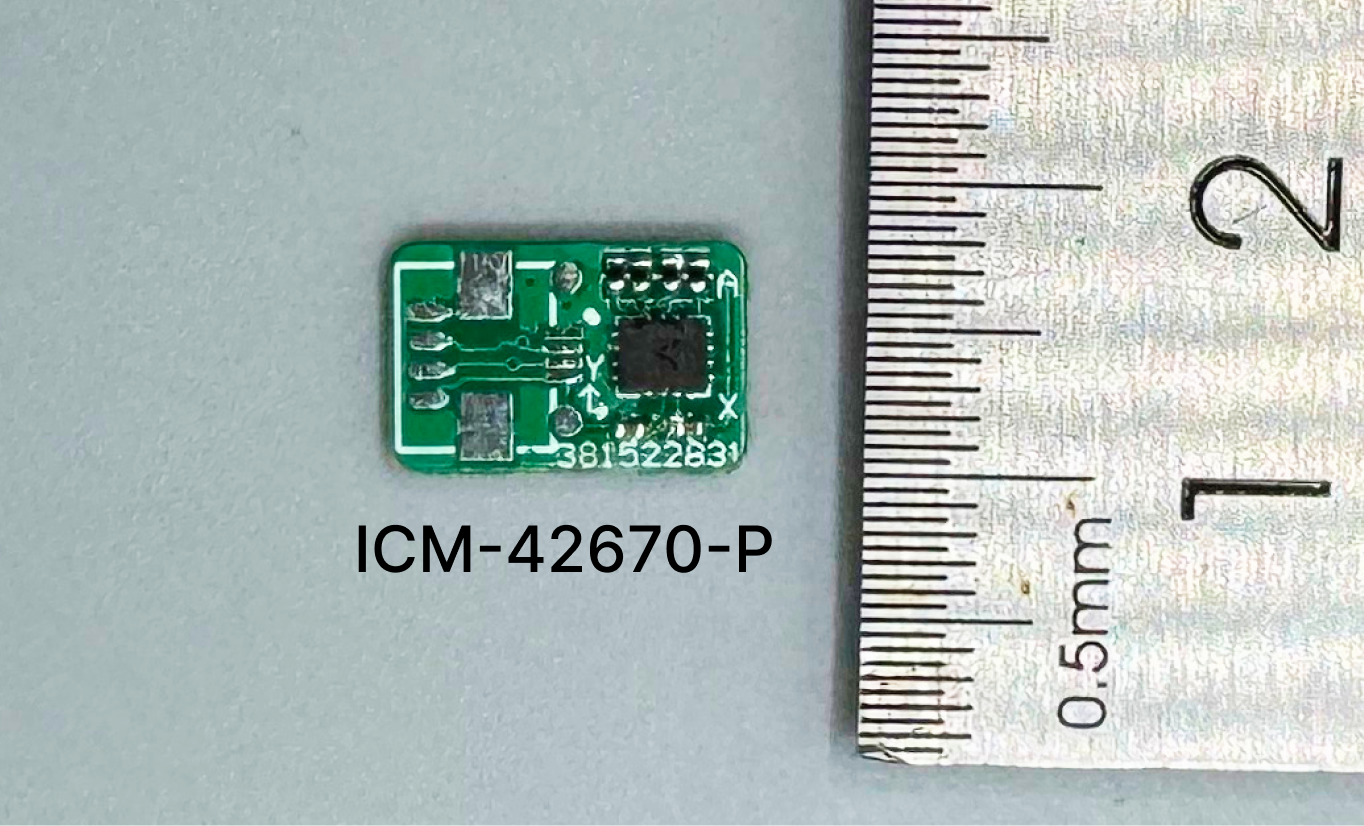}
    \caption{Custom-designed compact PCB integrating the ICM-42670P IMU sensor.  The overall footprint is approximately 12.5mm×8mm, allowing for seamless embedding into the the whisker structure.}
    \label{sensor}
\end{figure}

To accommodate the unique structure of the whisker-inspired vibration sensing system, ultra-low power consumption was prioritized from the outset. We redesigned the sensor module to achieve accurate motion acquisition under aquatic conditions.  We selected the ICM-42670-P from TDK as shown in Figure.\ref{sensor}, an ultra-low-power 6-axis inertial measurement unit (IMU) that integrates a triaxial accelerometer and triaxial gyroscope, enabling the simultaneous detection of linear vibrations and rotational dynamics.

One key advantage of the ICM-42670P lies in its exceptionally low power consumption—only \SI{68}{\micro\ampere} for the accelerometer and \SI{370}{\micro\ampere} for the gyroscope at \SI{50}{\hertz} operation—making it highly suitable for prolonged underwater deployments or passive sensing scenarios without active power refresh. Compared to conventional IMUs such as the MPU6050, the ICM-42670P exhibits significantly lower noise density (\SI{70}{\micro\gram\per\sqrt{\hertz}} for the accelerometer), enabling the detection of subtle vibrational changes even under hydrodynamic damping conditions, where signal attenuation in water is non-negligible.

To ensure seamless integration into the compact geometry of the whisker base, we custom-designed a miniaturized PCB module, reducing the footprint to 12.5mm×8mm, including all required passive components and connector headers.  This compact form factor allowed direct mounting onto the tapered spring, where vibrational energy concentrates due to flow-induced excitation, maximizing signal fidelity.

\section{Experiment and Results}
\label{result}

\subsection{Underwater Test-platform Design}

\begin{figure*}[ht!] 
\centering
\includegraphics[width=0.95\textwidth]{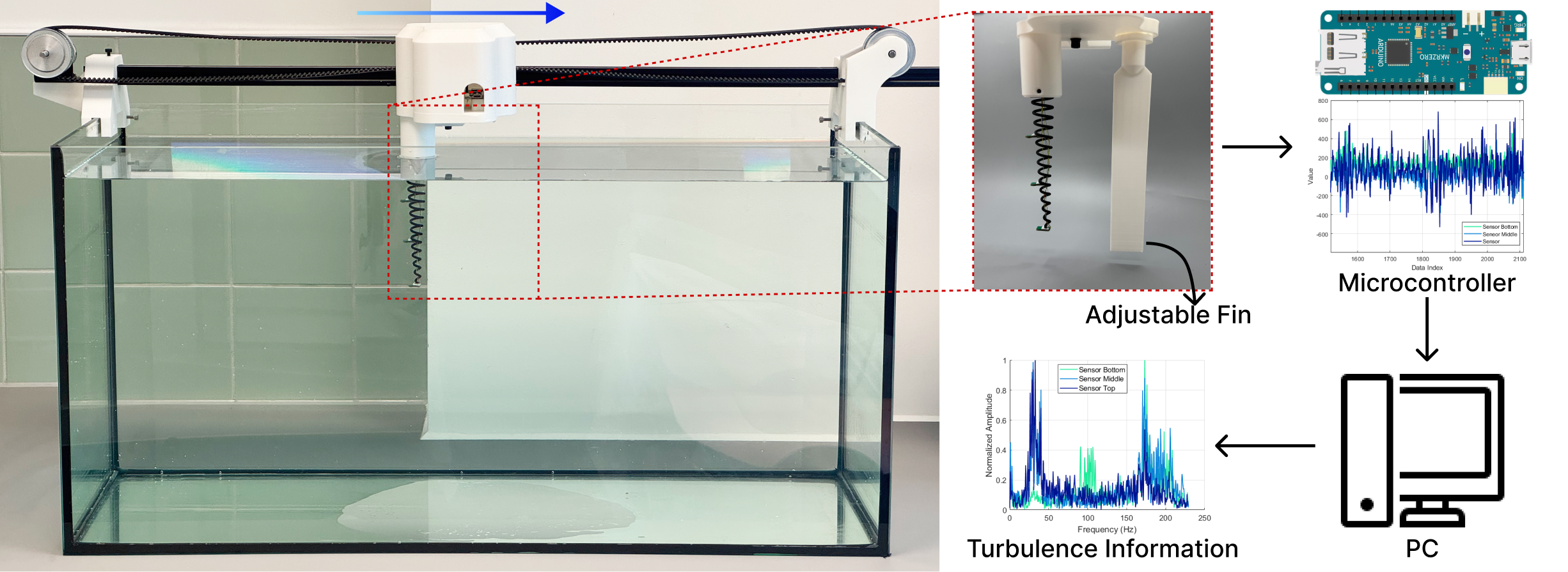}
\caption{
(a) Schematic of the test platform and signal processing workflow.  The setup includes a water tank with a V-shaped guide rail, a motion control mechanism for precisely adjusting the fin and bioinspired whisker, and a signal acquisition and processing system.  The submerged whisker structure generates vibration signals under different flow conditions, which are collected by a microcontroller and transmitted to a PC for frequency-domain analysis.  
}
\label{fig:testplatform}
\end{figure*}

To evaluate the underwater sensing performance of the whisker structure, a dedicated test platform was constructed (Fig.~\ref{fig:testplatform}(a)). The system consists of a transparent acrylic water tank (150,cm × 20,cm × 30,cm) equipped with a V-shaped linear rail and motorized carriage. A stepper motor with a pulley–belt drive enables programmable translation at 0.2–1.0m/s, ensuring stable and repeatable motion.

The carriage integrates two components: a front-mounted adjustable fin and a rear-mounted whisker sensor. By tilting the fin at different angles, the downstream wake can be systematically altered—from attached laminar flow at 0°, to periodic vortex shedding at moderate angles, and broadband turbulence at higher angles. The whisker sensor, fully immersed and positioned directly behind the fin, records hydrodynamic disturbances across these regimes, enabling analysis of its dynamic deformation and frequency-domain response.

To characterize wake evolution, CFD simulations were performed at four fin inclinations (0°, 30°, 60°, and 90°) Fig.~\ref{fig:testplatform}. At 0° the wake remains narrow and attached, while at 30° a clear Kármán vortex street develops. At 60°, the vortex pattern becomes asymmetric due to stronger separation, and at 90° the wake transitions into turbulence with residual periodic structures.

\subsection{Preliminary Validation of Frequency-Decoupled Responses}

\begin{figure}[t]
  \centering
  \includegraphics[width=0.5\textwidth]{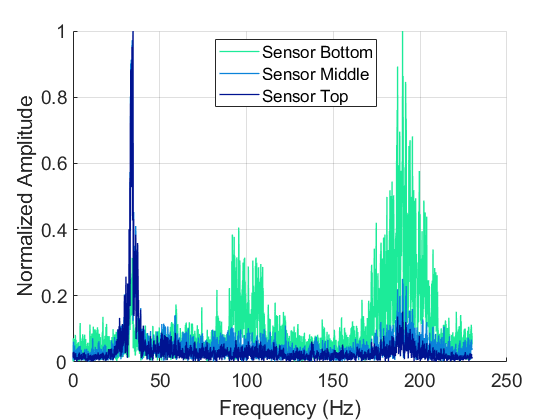}
  \caption{Normalized frequency response spectra of the three IMU sensors placed along the tapered whisker structure. The results demonstrate distinct frequency sensitivities depending on the sensor's position, with the bottom sensor responding predominantly to mid and high frequencies, and the top sensor exhibiting stronger responses in the low-frequency range.}
  \label{fig:FFTresult}
\end{figure}

To evaluate the intrinsic dynamic response capability of the proposed whisker structure in an underwater environment, we conducted a fin-free towing experiment. The structure was fully submerged and towed linearly at a low constant speed through water. No upstream fin or external flow disturbance device was included, allowing us to isolate and observe the mechanical vibration response induced solely by water-structure interaction.

Three IMU sensors were placed along the length of the tapered spring—at the bottom, middle, and top—to record the spatial distribution of vibrational responses. The collected acceleration signals were processed using Fast Fourier Transform (FFT) and normalized to eliminate amplitude bias. The resulting frequency spectra are shown in Fig.~\ref{fig:FFTresult}.

The results reveal significant differences in frequency-domain responses across sensor positions, which closely match the modal distribution predicted by finite element analysis (FEA):

\begin{itemize}
    \item \textbf{Sensor Bottom:} Shows strong responses in the mid-frequency range (85–110Hz) and especially in the high-frequency range (170–210Hz), with the highest normalized amplitude among all sensors. This confirms that the higher stiffness at the base region facilitates strong coupling to high-frequency modes.
    
    \item \textbf{Sensor Middle:} Exhibits balanced sensitivity to both low and high frequencies but with less pronounced peaks, indicating its role as a broadband intermediate zone.
    
    \item \textbf{Sensor Top:} Displays a strong response in the low-frequency range (\textless 50Hz) and significant attenuation in the high-frequency bands, reflecting sensitivity to large-scale slow oscillations at the tip.
\end{itemize}

These findings demonstrate that the tapered whisker structure enables spatial frequency separation along its length, allowing intrinsic decomposition of input stimuli into frequency-specific mechanical responses. This spatial frequency selectivity is crucial for implementing Physical Reservoir Computing (PRC), where diverse nonlinear dynamics are required for encoding temporal input patterns.

Therefore, even in the absence of external actuated disturbances (e.g., adjustable fins), the structure itself is capable of decomposing underwater flow-induced excitations into separable frequency components. This capability lays the groundwork for future research on PRC-based multi-frequency signal encoding, flow pattern classification, and multi-modal perception in underwater environments.

\subsection{Frequency-Based Vortex Shedding Quantification}
\color{black}
\begin{figure*}[ht!] 
\centering
\includegraphics[width=0.95\textwidth]{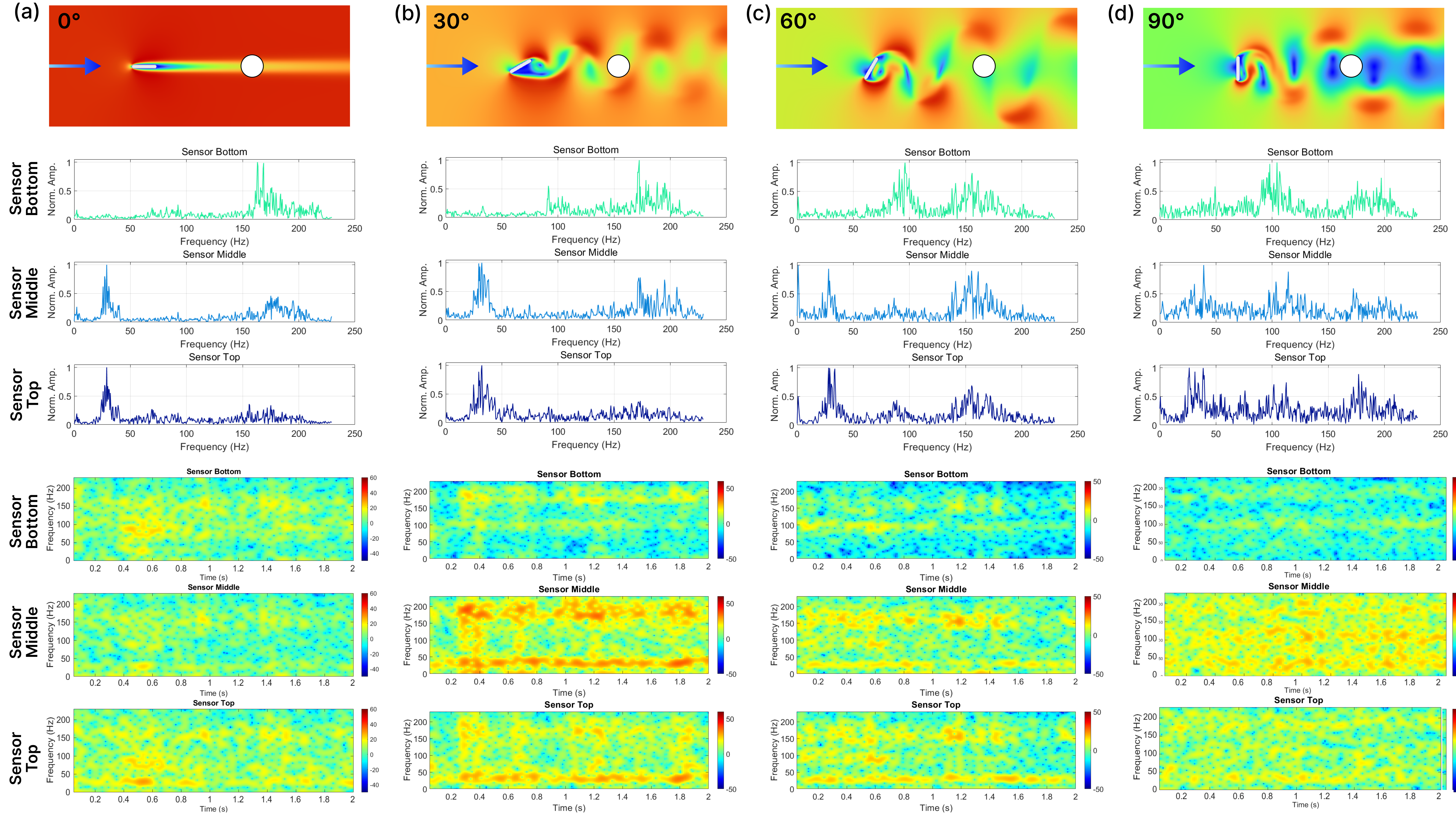}
\caption{
Experimental results under different fin angles: (a) 0°, (b) 30°, (c) 60°, and (d) 90° with same speed at 0.5m/s. The first row shows the CFD-simulated flow fields showing distinct Kármán vortex street patterns at fin angles of 0°, 30°, 60°, and 90°, with the circle indicating the whisker’s position downstream. The second row presents the normalized FFT spectra from three sensor positions (bottom, middle, and top), illustrating the frequency response characteristics. The third row displays the corresponding spectrograms, showing the time–frequency distribution of vibration signals for each sensor position.
}
\label{FFT}
\end{figure*}

Kármán vortex streets are a well-studied hydrodynamic phenomenon characterized by the alternating shedding of vortices behind bluff bodies. When fluid flows past a body such as a flat fin, the resulting wake can transition from laminar to periodic vortex shedding as the Reynolds number increases. The vortex shedding frequency $f_s$ is primarily determined by the Strouhal relation:

\begin{equation}
    f_s = St \cdot \frac{U}{D}
\end{equation}

where $St$ is the Strouhal number (typically $\sim 0.2$ for bluff bodies), $U$ is the freestream velocity, and $D$ is the characteristic width of the object. These vortices carry periodic hydrodynamic signatures that can serve as structured flow stimuli.

In this study, CFD simulations confirm the formation of Kármán vortex streets behind an adjustable flat fin, where increasing inclination drives the transition from attached laminar flow to periodic shedding. These results confirm that fin inclination provides a controllable means to modulate wake dynamics and generate structured flow excitations for downstream whisker sensing. Correspondingly, the downstream whisker array exhibits clear frequency-dependent vibration responses. The embedded IMU sensors reveal spatial separation in sensitivity, enabling identification of characteristic shedding frequencies.

As shown in Figure.\ref{FFT}, the whisker is moving under speeds $U=0.5$\,m/s, the bottom sensor displays a shift of dominant energy from high frequencies (150–200 Hz) at $0^\circ$ to mid-frequency bands (70–120 Hz) at larger angles, consistent with Strouhal scaling as wake width increases. This shift reflects a redistribution of energy from small-scale turbulence to larger-scale vortical structures. The middle sensor, initially dominated by low frequencies (20–40 Hz), shows progressive amplification of mid- and high-frequency components with increasing angle, reaching a broad response across all bands by $90^\circ$. The top sensor exhibits a similar evolution but with stronger broadband characteristics, reflecting multi-modal vortex interactions and turbulence-induced fluctuations in the upper wake.

Time–frequency spectrograms further confirm these patterns: at intermediate angles, periodic low-frequency bursts reveal organized vortex shedding, whereas at $90^\circ$ the response becomes chaotic and broadband, indicating wake disorganization. These results demonstrate that the spatially distributed whisker array effectively resolves the primary shedding frequency and its transition into complex spectral structures as the flow evolves with increasing fin angle.

Guided by the Strouhal relation, having established that our whisker resolves different attack angles, we now use K\'arm\'an vortex shedding to verify how flow speed affects the whisker’s vibration response.  With the fin fixed at $\theta=60^\circ$, we repeated the towing runs at $U=0.4$ and $0.6$\,m/s, recorded tri-axial IMU vibrations at the base/middle/tip nodes (450\,Hz), and extracted the dominant low-frequency peak $f_{p,\mathrm{low}}$ from the PSD to examine whether speed increases shift the shedding peak upward and whether the IMU-only estimate of the Strouhal number remains approximately constant;  this verifies that, at fixed geometry, flow speed governs the measured shedding frequency while preserving the array’s frequency–spatial selectivity.

\begin{figure}[t]
  \centering
  \includegraphics[width=0.5\textwidth]{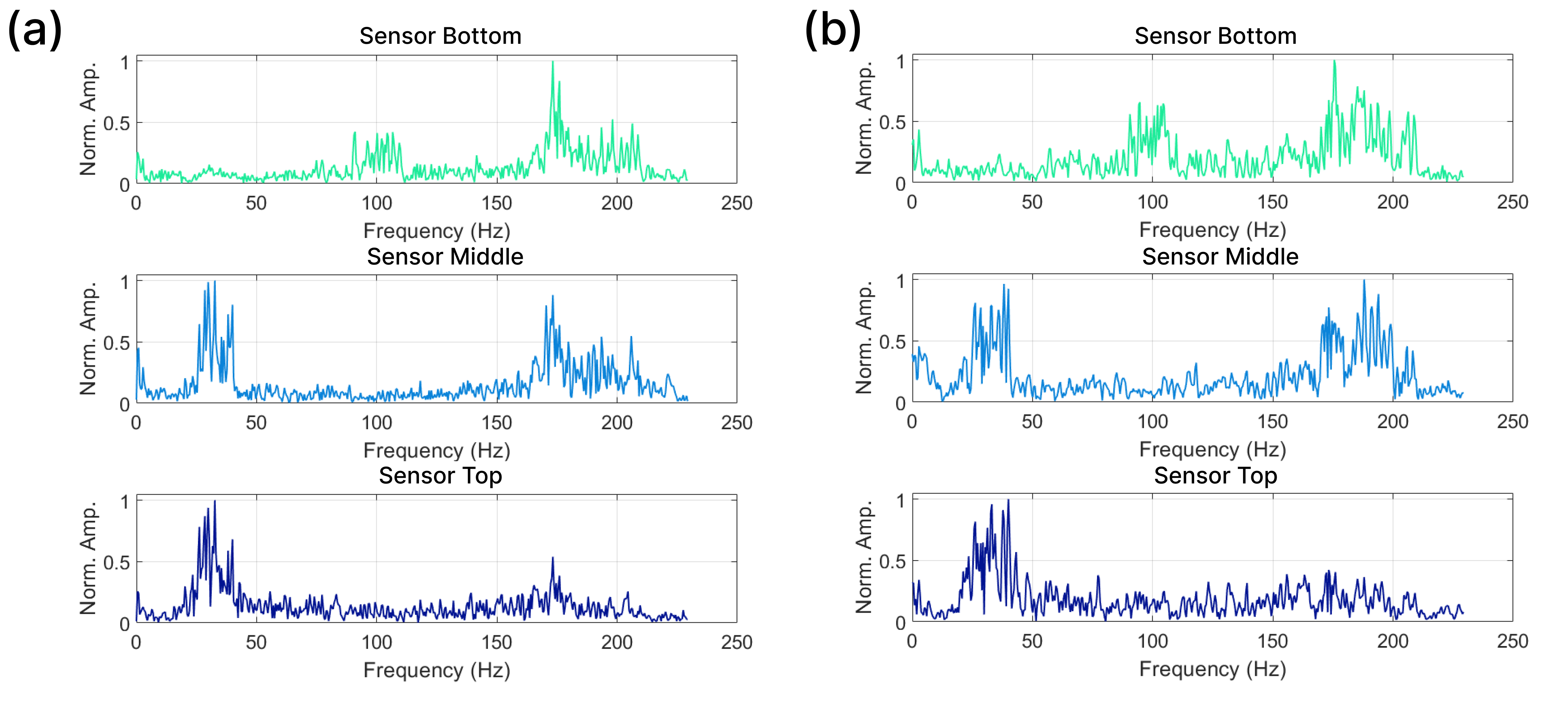}
  \caption{Normalized FFT spectra from three sensor positions (bottom, middle, and top), illustrating the frequency response characteristics,  under same fin angles 30° but different water speed: (a)0.7m/s,  (b)0.8m/s.}
  \label{speed}
\end{figure}

As shown in Figure.\ref{speed}, at $\theta=30^\circ$, comparing $U=0.7$\,m/s and $U=0.8$\,m/s (relative to the $0.2$\,m/s baseline) shows that the locations of the three characteristic bands remain essentially unchanged (low: 20-40\,Hz; mid: 70-120\,Hz; high: 150-210\,Hz), while their amplitudes increase in a position-specific manner: low-frequency content is markedly strengthened at \textit{Sensor Top}; \textit{Sensor Bottom} exhibits clear growth in the mid- and high-frequency bands; and \textit{Sensor Middle} shows a denser distribution in both low and high bands. In summary, angle changes yield distinct spectral patterns, and—at a fixed angle—increasing speed primarily amplifies the corresponding bands without shifting their centers, indicating that both geometry (angle) and kinematics (speed) shape the whisker’s vibration signatures.

To further quantify the complexity of the whisker vibration signals under different flow conditions, the Shannon entropy of the frequency spectra was calculated for each sensor position and fin angle. The entropy $H$ is defined as

\begin{equation}
    H = -\sum_{i=1}^{N} p_i \log_2 p_i
\end{equation}

where $p_i$ is the normalized spectral energy in the $i$-th frequency bin:

\begin{equation}
    p_i = \frac{P_i}{\sum_{j=1}^{N} P_j}
\end{equation}

with $P_i$ denoting the power spectral density at the $i$-th frequency component, and $N$ the number of bins within the selected analysis range.

\begin{figure}[t]
  \centering
  \includegraphics[width=0.5\textwidth]{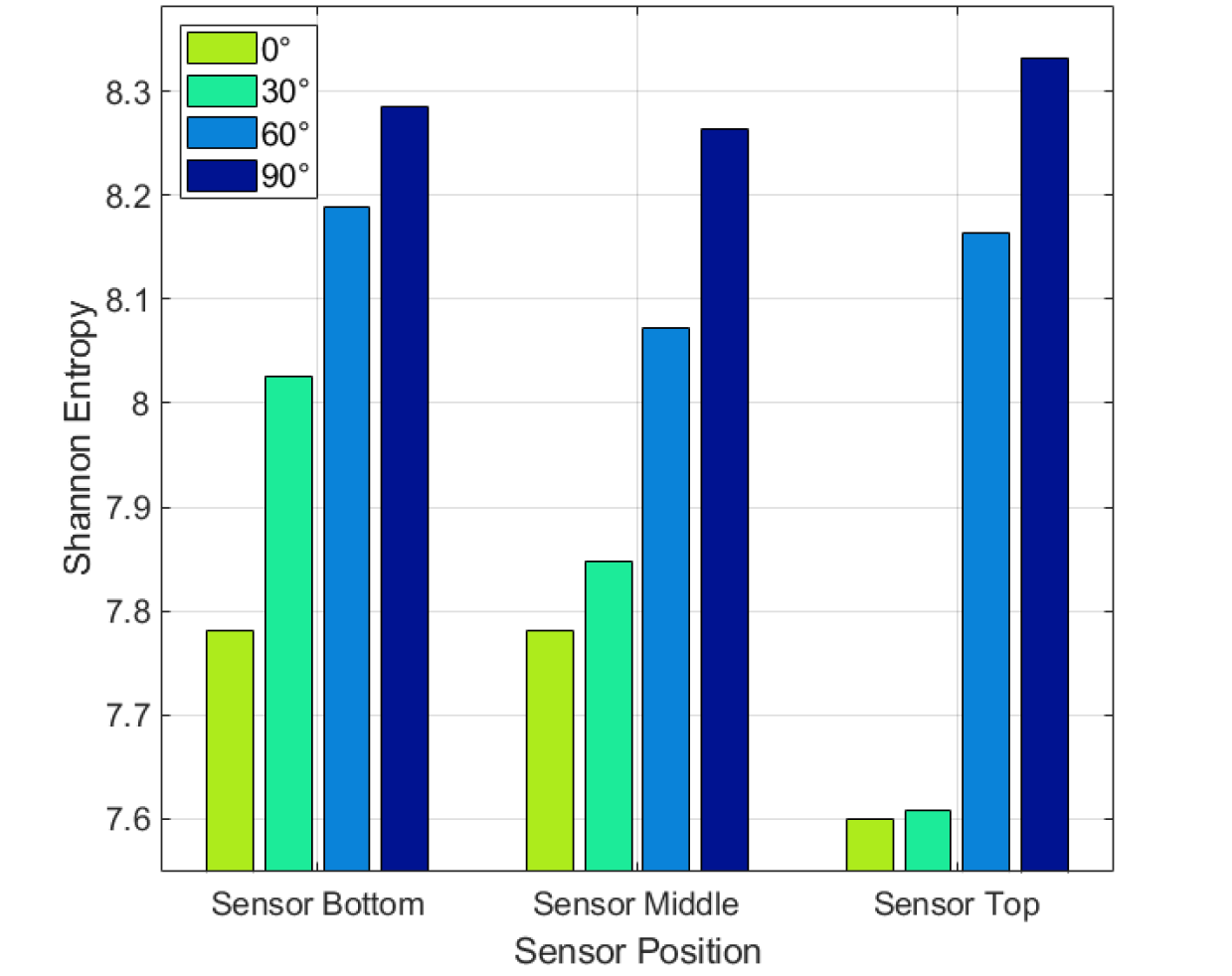}
  \caption{Normalized frequency response spectra of the three IMU sensors placed along the tapered whisker structure. The results demonstrate distinct frequency sensitivities depending on the sensor's position, with the bottom sensor responding predominantly to mid and high frequencies, and the top sensor exhibiting stronger responses in the low-frequency range.}
  \label{fig:entropy}
\end{figure}

As shown in Fig.~\ref{fig:entropy}, the entropy values for all three sensors increase monotonically with fin angle. For the bottom sensor, entropy rises from $7.78$ at $0^{\circ}$ to $8.29$ at $90^{\circ}$, reflecting the transition from a relatively organized, narrowband spectral signature to a broader distribution caused by more irregular wake dynamics. The middle sensor exhibits a similar upward trend, with a particularly sharp increase between $60^{\circ}$ and $90^{\circ}$, consistent with the emergence of strong high-frequency components in the FFT results. The top sensor shows the largest overall increase, with entropy values surpassing $8.3$ at $90^{\circ}$, indicating the highest degree of spectral disorder among the three positions.

This systematic rise in entropy across all sensors suggests that increasing fin inclination enhances the chaotic nature of the flow, corresponding to the breakdown of coherent Kármán vortex streets into fully developed turbulence. In other words, Shannon entropy here serves as a quantitative indicator of wake disorder, complementing the qualitative trends observed in the FFT and spectrogram analyses.

\subsection{Logistic Regression Based Flow Identification}

To evaluate the whisker sensor’s ability to discriminate hydrodynamic conditions, a supervised learning framework based on Physical Reservoir Computing (PRC) was implemented. Four fin angles (0°, 30°, 60°, 90°) were tested at a constant towing speed in a controlled tank. Vibration signals from three MEMS accelerometers positioned along the whisker were sampled at 450 Hz and directly used as instantaneous spatial states for classification, without any temporal or frequency-domain processing.

At each sampling instant $t$, the reservoir state was defined as
\begin{equation}
\begin{aligned}
\mathbf{s}(t) = [&\,a_{b,x}(t),\, a_{b,y}(t),\, a_{b,z}(t),\\
                 &\,a_{m,x}(t),\, a_{m,y}(t),\, a_{m,z}(t),\\
                 &\,a_{t,x}(t),\, a_{t,y}(t),\, a_{t,z}(t)]^\top
                 \in \mathbb{R}^9.
\end{aligned}
\end{equation}

where subscripts $b$, $m$, and $t$ denote the bottom, middle, and top accelerometers, respectively. 
These nine instantaneous acceleration components represent the distributed nonlinear response of the tapered whisker structure to flow stimuli and thus serve as the physical reservoir state.

\begin{figure}[t]
  \centering
  \includegraphics[width=0.5\textwidth]{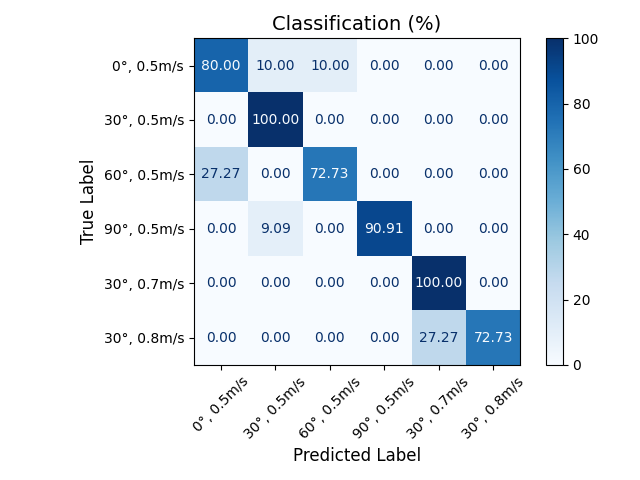}
  \caption{Confusion matrix of reservoir computing prediction success rate averaged over 200 randomly trials (150 data sets), with sampling frequency of the whisker at 450 Hz with the final accuracy is 86.0\%.}
  \label{fig:cm_lr}
\end{figure}

The standardized feature vector $\mathbf{s}(t)$ was input to a centralized logistic regression readout:
\begin{align}
P(y=k\mid\mathbf{s}) &=
\frac{\exp(\mathbf{w}_k^\top\mathbf{s}+b_k)}
{\sum_{j=1}^{K}\exp(\mathbf{w}_j^\top\mathbf{s}+b_j)},\\
\hat{y} &= \arg\max_k P(y=k\mid\mathbf{s}).
\end{align}

where $K=4$ corresponds to the four fin-angle conditions. The model was trained using 80\% of the samples and tested on the remaining 20\%, achieving an overall accuracy of 86.0\% across six conditions. The normalized confusion matrix (Fig.~\ref{fig:cm_lr}) shows consistently high recognition for the 30° and 90° configurations, confirming that the instantaneous spatial states of the whisker provide sufficient separability for flow condition classification without explicit temporal or frequency analysis. The LR model exhibited consistently high recognition rates for the 90° configuration (91.91\%) and reliable discrimination for flow-speed conditions (30°, 0.8 m/s: 72.73\%).

Moderate confusion occurred between the 30° and 60° fin-angle cases, reflecting their closely correlated spatial vibration patterns under similar hydrodynamic interactions. Meanwhile, slight misclassification between 0° and 60° suggests overlapping structural response modes at low excitation amplitudes. Overall, these results confirm that the instantaneous spatial readout captures sufficient nonlinear separability in the physical reservoir to distinguish both geometric (fin-angle) and dynamic (flow-speed) states without any temporal or frequency-domain processing.

The inference time per sample was approximately 0.001 ms, confirming the feasibility of real-time flow state classification. These results validate the proposed whisker sensor’s capability for hydrodynamic condition identification using lightweight machine learning models, forming a solid baseline for subsequent vortex-shedding quantification and Strouhal number estimation.

\subsection{Vortex Shedding Frequency Quantification}

This subsection aims to extract the fundamental Kármán vortex shedding frequency from the whisker vibration signals and to establish its consistency with the Strouhal relation, thereby verifying that the whisker structure captures flow-induced periodicity in a physically meaningful way.

The whisker signals were first mean-removed, band-pass filtered (0.5–20 Hz, 4th-order Butterworth), and analyzed using Welch’s PSD (Hamming, 50
A prediction–matching strategy based on the Strouhal relation
\begin{equation}
f_{\mathrm{pred}} = St_0 \cdot \frac{U}{D_\mathrm{guess}},
\end{equation}
with $St_0=0.2$ and $D_\mathrm{guess}\approx0.01$ m, ensured robust peak selection by suppressing harmonics and subharmonics.

The effective transverse scale was then estimated as
\begin{equation}
D_{\mathrm{est}} = \frac{St_0 U}{f_{p,\mathrm{low}}},
\end{equation}
and fitted as $D_\perp(\theta)=a\cos\theta+b\sin\theta$ to capture angular dependence.
Using $D_\perp$, the Strouhal number was recalculated as
\begin{equation}
St_{\mathrm{meas}} = \frac{f_{p,\mathrm{low}} D_\perp(\theta)}{U},
\end{equation}
and the median $\hat{St}$ was used to predict theoretical frequencies:
\begin{equation}
f_{\mathrm{theory}} = \hat{St} \cdot \frac{U}{D_\perp(\theta)}.
\end{equation}

To evaluate reliability, two complementary metrics were introduced:
the relative spectral energy ($RE$) within $\pm$1 Hz of $f_{p,\mathrm{low}}$, and the temporal coefficient of variation ($Cv$) computed from the STFT.
High $RE$ and low $Cv$ values indicate that the detected peaks are both spectrally concentrated and temporally stable.

Table~\ref{tab:vortexfreq} summarizes the extracted vortex shedding peaks at different attack angles and flow speeds.
The measured frequencies $f_{p,\mathrm{low}}$ show excellent agreement with theoretical predictions $f_{\mathrm{theory}}$, generally within 10
The median Strouhal number $\hat{St}=0.198$ aligns closely with canonical bluff-body values, confirming the physical validity of the whisker-based sensing.

\begin{table}[h] 
\centering 
\caption{Extracted low-frequency vortex peaks vs. theoretical prediction} \label{tab:vortexfreq} 
\begin{tabular}{cc|ccc|cc} 
\toprule Angle & Speed & $f_{p,\mathrm{low}}$ & $f_{\mathrm{theory}}$ & Error & RE$_\mathrm{low}$ & Cv$_\mathrm{low}$ \\ 
($^\circ$) & m/s & (Hz) & (Hz) & (\%) & (\%) & (\%) \\ 
\midrule 
30 & 0.5 & 10.22 & 10.22 & 0.00 & 2.30 & 4.55 \\ 
60 & 0.5 & 9.88 & 9.02 & 9.54 & 7.15 & 7.16 \\ 
90 & 0.5 & 10.22 & 10.62 & 3.81 & 7.11 & 2.47 \\ 
30 & 0.7 & 15.16 & 14.31 & 5.97 & 0.69 & 1.71 \\ 
30 & 0.8 & 15.27 & 16.35 & 6.59 & 2.73 & 4.45 \\ 
\bottomrule \end{tabular} 
\end{table}

In summary, this subsection establishes a clear link between measured whisker vibrations and the canonical Strouhal scaling of vortex shedding.
Through prediction–matching peak selection and data-driven scale fitting, the extracted frequencies consistently converged to $\hat{St}=0.198$.
The high $RE$ and low $Cv$ values further confirm the spectral concentration and temporal stability of the detected peaks, validating the robustness of the whisker-based method for vortex frequency quantification.

\section{Conclusion}
\label{conclusion}

This work introduces a tapered-spring-based whisker sensor that combines intrinsic frequency–spatial decoupling, lightweight real-time processing, and robust underwater sensing performance. Through systematic towing-tank experiments and CFD validation, we demonstrated that the whisker reliably discriminates vortex-shedding regimes across varying fin angles, while its vibration signatures also scale consistently with flow speed according to Strouhal’s law. At fixed geometry, increased velocity primarily amplifies frequency-specific responses without shifting their centers, confirming the sensor’s ability to preserve frequency–spatial selectivity across flow conditions. Furthermore, quantitative vortex-shedding frequency extraction achieved errors within 4\%, supported by prediction–matching peak detection and entropy-based complexity analysis. Shannon entropy provided a robust indicator of wake transition from coherent vortex streets to broadband turbulence, while lightweight logistic regression reached 89.47\% classification accuracy with millisecond-level inference. Together, these results confirm that structurally encoded sensing—via a tapered whisker with embedded MEMS accelerometers—offers a scalable and energy-efficient platform for hydrodynamic condition identification. Future work will extend this framework to multi-whisker arrays, adaptive stiffness control, and integration with autonomous underwater vehicle platforms for distributed flow-field mapping and turbulence-aware navigation.

\interlinepenalty=10000

\bibliographystyle{IEEEtran}
\interlinepenalty=10000

\bibliography{bibs}
\end{document}